\documentclass[journal]{IEEEtran}
\usepackage{amsmath,amsfonts}
\usepackage{algorithmic}
\usepackage{algorithm}
\usepackage{array}
\usepackage{colortbl}
\usepackage{enumitem}
\usepackage[caption=false,font=normalsize,labelfont=sf,textfont=sf]{subfig}
\usepackage{textcomp}
\usepackage{stfloats}
\usepackage{url}
\usepackage{verbatim}
\usepackage{graphicx}
\usepackage{cite}
\usepackage{orcidlink}
\usepackage[nolist]{acronym}
\usepackage{adjustbox}
\usepackage{multirow}
\definecolor{red_cool}{rgb}{0.5, 0.0, 0.0}
\definecolor{Gray}{gray}{0.85}
\definecolor{green_cool}{rgb}{0.0, 0.5, 0.0}

\begin{acronym}
\acro{BSMD-STD}{binaural signal magnitude difference standard deviation}
\acro{MAE}{mean absolute error}
\acro{CNN}{convolutional neural network}
\acro{CRNN}{convolutional recurrent neural network}
\acro{DoA}{direction of arrival}
\acro{DNN}{deep neural network}
\acro{DRR}{direct-to-reverberant energy ratio}
\acro{ELU}{exponential linear unit}
\acro{GMM}{Gaussian mixture model}
\acro{GRU}{gated recurrent unit}
\acro{IID}{interchannel intensity difference}
\acro{ITD}{interchannel time difference}
\acro{ILD}{interchannel level difference}
\acro{KNN}{K-nearest neighbours}
\acro{LDA}{linear discriminative analysis}
\acro{LSTM}{long short-term memory}
\acro{LPC}{linear predictive coding}
\acro{MSC}{magnitude squared coherence}
\acro{MSE}{mean squared error}
\acro{MEMS}{micro-electromechanical systems}
\acro{RIR}{room impulse response}
\acro{RNN}{recurrent neural network}
\acro{SAM}{spectrogram-aware attention module}
\acro{SDE}{source distance estimation}
\acro{SED}{sound event detection}
\acro{SELD}{sound event localization and detection}
\acro{SNR}{signal-to-noise ratio}
\acro{SVM}{support vector machine}
\acro{STFT}{short-time Fourier transform}
\acro{SDE}{source distance estimation}
\acro{DOAE}{direction of arrival estimation}
\end{acronym}

\begin{document}

\title{Speaker Distance Estimation in Enclosures from Single-Channel Audio}

\author{Michael~Neri~\orcidlink{0000-0002-6212-9139},~\IEEEmembership{Graduate Student~IEEE,} 
        Archontis~Politis~\orcidlink{0000-0002-0595-2356},~\IEEEmembership{Member~IEEE,}
        Daniel~Krause~\orcidlink{0000-0003-2704-636X},~\IEEEmembership{Graduate Student~IEEE,} 
        Marco~Carli~\orcidlink{0000-0002-7489-3767},~\IEEEmembership{Senior~IEEE,}
        and~Tuomas~Virtanen~\orcidlink{0000-0002-4604-9729},~\IEEEmembership{Fellow~IEEE}
\thanks{
M. Neri and M. Carli are with the Department of Industrial, Electronic, and Mechanical Engineering, Roma Tre University, Rome, Italy. (e-mail: \href{mailto:michael.neri@uniroma3.it}{michael.neri@uniroma3.it}, \href{mailto:marco.carli@uniroma3.it}{marco.carli@uniroma3.it}) \\
A. Politis, D. Krause, and T. Virtanen are with the Faculty of Information Technology and Communication Sciences, Tampere University, Tampere, Finland. (e-mail: \href{mailto:archontis.politis@tuni.fi}{archontis.politis@tuni.fi}, \href{mailto:daniel.krause@tuni.fi}{daniel.krause@tuni.fi}, \href{mailto:tuomas.virtanen@tuni.fi}{tuomas.virtanen@tuni.fi}).
}
}

\markboth{IEEE/ACM Transactions on Audio, Speech, and Language Processing,~Vol.~XX, No.~X, March~2024}%
{Neri \MakeLowercase{\textit{et al.}}: Speaker Distance Estimation in Enclosures from Single-Channel Audio}


\maketitle

\begin{abstract}
Distance estimation from audio plays a crucial role in various applications, such as acoustic scene analysis, sound source localization, and room modeling. Most studies predominantly center on employing a classification approach, where distances are discretized into distinct categories, enabling smoother model training and achieving higher accuracy but imposing restrictions on the precision of the obtained sound source position. Towards this direction, in this paper we propose a novel approach for continuous distance estimation from audio signals using a convolutional recurrent neural network with an attention module. The attention mechanism enables the model to focus on relevant temporal and spectral features, enhancing its ability to capture fine-grained distance-related information. To evaluate the effectiveness of our proposed method, we conduct extensive experiments using audio recordings in controlled environments with three levels of realism (synthetic room impulse response, measured response with convolved speech, and real recordings) on four datasets (our synthetic dataset, QMULTIMIT, VoiceHome-2, and STARSS23). Experimental results show that the model achieves an absolute error of 0.11 meters in a noiseless synthetic scenario. Moreover, the results showed an absolute error of about 1.30 meters in the hybrid scenario. The algorithm's performance in the real scenario, where unpredictable environmental factors and noise are prevalent, yields an absolute error of approximately 0.50 meters. For reproducible research purposes we make model, code, and synthetic datasets available at \url{https://github.com/michaelneri/audio-distance-estimation}.
\end{abstract}


\section{Introduction}

\IEEEPARstart{S}OURCE distance estimation (SDE) refers to the task of estimating the interspace between a microphone and a sound source. It is very often performed in conjunction with \ac{DoA} estimation, in which only the direction information about the source position is obtained. Both tasks are useful in many practical applications, including increasing the robustness of automatic speech recognition~\cite{wolfel2009distant} by enhancing the performance of acoustic echo cancellers \cite{Bekrani_TASL_2011} and autonomous robotics~\cite{Berglund2005,Rodemann2010}. Despite both \ac{DoA} and source distance being estimated using multi-channel audio in most practical scenarios, the latter has been largely under-researched~\cite{BrendelDistance}. Firstly, source distance estimation is widely regarded a more difficult task due to distance cues vanishing with the increased space between the sound source and the receiver. Secondly, \ac{DoA} offers sufficient information in many downstream spatial filtering tasks. However, many applications such as source separation, acoustic monitoring, and context-aware devices, would still benefit from full information about the sound source position, hence the need for further investigations on \ac{SDE}.

Most methods for both DOA and distance estimation rely on arrays with more than two microphones~\cite{DiBiase2000}. Multichannel data allows for exploiting spatial cues such as \acp{ITD} and \acp{ILD} to provide information for efficient \ac{DoA} estimation, positively affecting distance estimation as well~\cite{Rodemann2010}. However, using multiple microphones poses certain limitations in terms of budget and physical portability. To tackle this problem, some studies investigated using binaural recordings for that purpose, allowing for decreasing the number of channels to two by exploiting the human hearing cues~\cite{Yiwere2017DistanceEA, krause2021joint}. However, the simplest scenario of estimating distance from a single microphone has been largely under-researched~\cite{Patterson_2022_Interspeech}. Moreover, the vast majority of studies focus on a classification approach, in which the distance is discretized into a set of disjunctive categories, e.g., ``far" and ``near", allowing for easier model training and a higher accuracy~\cite{Georganti_2013_TASLP, Georganti_TASLP_2011}. However, using pre-defined categories does not allow for continuous estimation, which puts limits on the precision of the obtained sound source position.

\begin{figure*}[ht]
    \centering
\centerline{\includegraphics[width=1\linewidth]{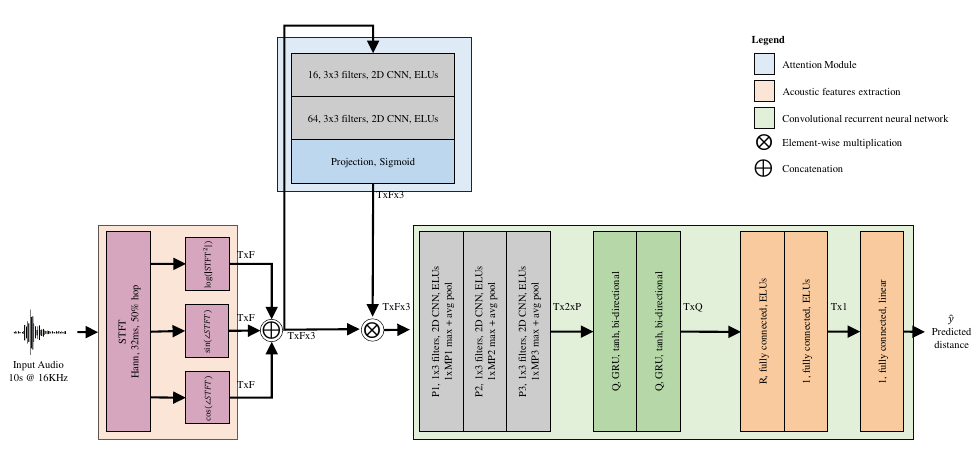}}
    \caption{Proposed architecture for speaker distance estimation. First, acoustic features are extracted from the single-channel audio. In more detail, $3$ maps (magnitude of the \ac{STFT}, sinus, and cosinus of the \ac{STFT} phase) are obtained with shape $T \times F$, where $T$ and $F$ are the time and frequency bins, respectively. Then, the maps are stacked along the channel dimension resulting in a feature tensor of size $T \times F \times 3$. To highlight the feature regions that are most informative for distance estimation, an attention map is learned from the three-channel tensor, which is then element-wise multiplied with the input feature tensor. The output is further processed by the convolutional layers with $P_i$ $1 \times 3$ kernels, also denoted as \textit{frequency kernels}, yielding a $T \times 2 \times P$ tensor that is arranged in a $T \times Q$ matrix, where $Q = 2P$. Subsequently, the resulting matrix is analyzed by two \ac{GRU} layers with $Q$ neurons to model temporal patterns. Finally, the output from recurrent layers $T \times Q$ is fed to three fully connected layers with $R$, $1$, and $1$ neurons respectively to map the features to the predicted distance $\hat{y}$.}
    \label{fig:model}
\end{figure*}

Towards this direction, in this work, we propose several novel solutions to tackle the problem of source distance estimation. Firstly, we define the task as a regression problem, differently from most state-of-the-art works that focus on classification-based methods. We propose a novel approach to distance estimation from single-channel audio signals in reverberant environments, overcoming the need for complex microphone arrays. In more detail, the proposed model is a \ac{CRNN} with an attention module, which is responsible for learning a time-frequency attention map. By doing so, it is possible to emphasize magnitude- and phase-related features that are the most informative for sound source distance estimation. The effectiveness of our approach is extensively tested for numerous acoustic scenarios, obtained by simulations with randomized configurations of room shapes, materials, and locations of the microphone and the speaker. In addition, tests have been carried out on real reverberant speech recordings, captured directly or emulated with real \acp{RIR}.

The remainder of the manuscript is organized as follows. Section~\ref{sec:related} provides a summary of the state-of-the-art. Section~\ref{sec:method} describes the proposed methods, whereas the performance evaluations are in Section~\ref{sec:perf}. Section~\ref{sec:results} details the experimental results of the proposed approach on three acoustic scenarios. Finally, Section~\ref{sec:discussion} includes an overall discussion regarding the work, and Section~\ref{sec:conclusion} draws the conclusions. 

\section{Related Works}\label{sec:related}

\ac{SDE} involves determining the distance between a sound source and the receiver. When compared to \ac{DoA} estimation, \ac{SDE} is an area that has received significantly less attention and is generally considered more challenging. This is primarily due to the fact that the accuracy of distance estimation declines rapidly for small-sized arrays commonly used in practice even for relatively short distances from the center of the array (up to $3$-$4$~m). Several factors contribute to this phenomenon, including: a) the decrease in \ac{DRR} and \acp{SNR} as the source distance increases, b) the reduction in inter-channel level differences and constant inter-channel time differences as the source transitions from a spherical wave to a plane wave captured by the array.

The majority of studies related to \ac{SDE} show results in conjunction with the \ac{DoA} estimation task. Extensive research has been conducted on this subject for various acoustic systems that commonly use distributed microphone arrays. These systems encompass a range of setups, such as intelligent loudspeakers~\cite{nielsen2018}, spherical microphones~\cite{Gontmacher2012}, triangular configurations~\cite{gabriel2019}, and arrays of acoustic sensors~\cite{Hwang:17}. Simpler audio formats including binaural recordings have been investigated to a much lesser extent, including few studies with classical machine learning methods~\cite{Rodemann2010, GHAMDAN201787} and very limited research related to deep learning~\cite{Yiwere2017DistanceEA, krause2021joint}.

Regarding \ac{SDE} modeling in isolation, most of the research has been focused on parametric approaches and manually crafted features. These methods often utilize information such as the \ac{DRR}~\cite{Lu_2010_TASLP}, \ac{RIR}~\cite{Samarasinghe_2014_ISCCSP}, or signal statistics and binaural cues such as the \ac{IID}~\cite{Rodemann2010}. In some cases, classical machine learning techniques have been employed to leverage statistical features. For instance, a study by Brendel \textit{et al.} estimated the coherent-to-diffuse power ratio to determine the source-microphone distance via \acp{GMM}~\cite{BrendelDistance}. Vesa utilized \acp{GMM} trained with \ac{MSC} features to incorporate information about channel correlation~\cite{Vesa_2007_WASPAA, Vesa_2009_TASLP}. In~\cite{Zhagyparova_2021_TENSYMP}, the authors used \ac{MSC} on top of other features to train classifiers with methods such as \ac{KNN} or \ac{LDA}. Georganti \textit{et al.} introduced the \ac{BSMD-STD} and trained \acp{GMM} and \acp{SVM} using this feature~\cite{Georganti2013}. Most of these methods rely on compound algorithms that require careful tuning to adapt to varying acoustic conditions.

Until now, the exploration of source distance estimation using \acp{DNN} has been quite limited. Yiwere \textit{et al.} employed an approach inspired by image classification, utilizing \acp{CRNN} trained on log-mel spectrograms to classify three different distances in three distinct rooms~\cite{Yiwere_2019_Sensor}. Although the models demonstrated promising outcomes for data within the same environment, their performance significantly deteriorated when dealing with recordings from different rooms. In another endeavor, Sobhdel \textit{et al.} introduced relation networks to address this challenge through few-shot learning, which exhibited enhancements over conventional \acp{CNN}~\cite{Sobhdel_2021_ARXIV}. Both studies conducted tests within a limited range of specific distances, encompassing a close proximity of up to 3-4 meters at most. In~\cite{krause2021joint}, the authors conducted experiments for data covering distances for up to 8~m, however the model was classifying them into two binary classes denoted as ``far" and ``near".

Additionally, only a few works have addressed the topic of speaker distance estimation using single-channel audio. One of the first works employed low-level features such as \ac{LPC}, skewness, and kurtosis of the spectrum to classify the distance of a speaker~\cite{Georganti_TASLP_2011}. Venkatesan \textit{et al.} proposed both monaural and binaural features to train \acp{GMM} and \acp{SVM}~\cite{Venkatesan_2020_CirSys}. Regarding \ac{DNN} approaches, Patterson \textit{et al.} classified ``far" and ``near" speech in order to perform sound source separation from single-channel audios~\cite{Patterson_2022_Interspeech}. 

To the best of our knowledge, single-channel source distance estimation has been scarcely addressed as a regression problem, prioritizing classification approaches to ease model training. In addition, there are very few studies investigating the use of \acp{DNN} in this task. For these reasons, a learning-based approach for continuous estimation of the distance of the speaker is proposed. A first step towards continuous sound source distance estimation occurs in our preliminary study~\cite{Neri_2023_WASPAA} where a \ac{CRNN} was defined for estimating static speaker distance in simulated reverberant environments from a single omnidirectional microphone. However, that study was evaluated only on simulations, while in this work various degrees of realism are investigated, from simulated \acp{RIR}, to synthetic data with measured \acp{RIR}, to fully real recordings with distance-annotated sources. Hence, the potential of the method in a real-world scenario is demonstrated. In addition, the preliminary study was based on a simpler architecture without investigation on what architectural components contributed the most to the \ac{SDE}, while here the architecture is refined and enhanced, with better overall performance, and specific choices investigated in an ablation study.

To cope with these limitations, the contributions of this work are as follows
\begin{itemize}
    \item a major improvement of the results of the learning-based approach, i.e., a \ac{CRNN}, proposed in our preliminary study~\cite{Neri_2023_WASPAA} that simultaneously provides temporal frame-wise and utterance-wise distance estimation of the static audio source. In addition, an in-depth study regarding the model architecture is detailed;
    \item definition of an attention module that estimates the most significant time-frequency patterns from the input features for speaker distance estimation;
    \item experiments have been conducted on synthetic data, both in noiseless and noisy scenarios, to analyze the response of the proposed approach in controlled environments. Further tests on the \ac{CRNN} have been conducted on a constructed hybrid dataset, i.e., measured \acp{RIR} convolved with anechoic speeches, and two real recording datasets, demonstrating the generalization capabilities of the proposed approach.
\end{itemize}

\section{Proposed method}\label{sec:method}
In this section, a description of the acoustic features for the source distance estimation is provided. To process temporal, spatial, and spectral characteristics of these features, a \ac{CRNN} has been employed for the experiments. This type of model has shown good results in many studies for \ac{SELD} tasks~\cite{Adavanne_2019_JSTSP, Adavanne_2019_DCASE}. In addition, an attention module is introduced to learn an attention map on the time-frequency audio representation. The overall architecture is depicted in Figure~\ref{fig:model}.

\subsection{Acoustic features extraction}
All the operations on the audio files are performed at $16$~kHz. The selection of this sampling frequency is because the speech spectrum is mostly contained in the range $0$-$8$~kHz~\cite{Byrne_BJA_1977}. In addition, a lower frequency yields a lower number of samples, reducing the computational complexity of feature extraction and distance estimation. Initially, a pre-processing stage is employed to extract the complex \ac{STFT} $\mathrm{STFT}\{\mathbf{x}\} \in \mathbb{C}^{T \times F}$ from the single-channel audio signal $\mathbf{x} \in \mathbb{R}^{1 \times L}$, where $T$ is the number of time frames, $F$ the number of frequency bins, and $L$ the number of samples. This transformation is computed using a Hann window of length $32$~ms with $50\%$ overlap. Subsequently, the magnitude ($|\mathrm{STFT}\{\mathbf{x}\}| \in \mathbb{R}^{T \times F})$ and phase ($\angle\mathrm{STFT}\{\mathbf{x}\} \in \mathbb{R}^{T \times F})$ components of the \ac{STFT} are computed from the complex matrix. 

Sinus and cosinus maps of the phase spectrogram are computed by applying $\mathrm{sin}(\cdot)$ and $\mathrm{cos(\cdot)}$ functions element-wise, since the features provide a smoother continuous representation of the raw phase information. The concept of utilizing the phase spectrogram has been adopted from contemporary research on multichannel source separation~\cite{Sudo_2021_SII}, learning-based localization~\cite{Manaperi_2022_TASLP}, and speech enhancement~\cite{Wang_2020_TASLP} as phase information contains cues regarding the acoustic properties of the environment in which the sound propagates~\cite{Pandey_2019_ICASSP}. Tests conducted using the raw complex spectrogram in our scenario, i.e., two separate branches that processed real and imaginary parts, yielded unsatisfactory training performance. 

Finally, the magnitude of the \ac{STFT} and the sinus and cosinus maps are stacked into a $T \times F \times 3$ tensor. This representation is then fed into the attention module and the convolutional layers for further processing and analysis.

\subsection{Attention Module}
One of the main contributions of this work is the definition of an attention module which computes an attention map $H \in \mathbb{R}^{+T \times F \times 3}$ from the audio features. The objective of this learned matrix is to emphasize the regions of the features that are most informative for the estimation of the distance. Specifically, this module is the function $\mathrm{f_{\mathrm{ATT}}}:\mathbb{R}^{T \times F \times 3} \rightarrow \mathbb{R}^{{+}^{T \times F \times 3}}$. Its structure is composed of $2$ convolutional blocks, having $16$ and $64$ $3 \times 3$ filters, respectively. Then, a $1 \times 1$ convolutional layer with three filters, followed by a sigmoid activation, is used to map the features to yield the $T \times F \times 3$ attention map. Finally, the output acoustic features $\Tilde{X} \in \mathbb{R}^{T \times F \times 3}$ are obtained by element-wise multiplication ($\otimes$) between the input acoustic features and the attention map as 
\begin{equation}
    \Tilde{X} = f_{\mathrm{ATT}}(X) \otimes X.
\end{equation}

\begin{figure*}[ht]
    \centering
\centerline{\includegraphics[width=1.10\linewidth]{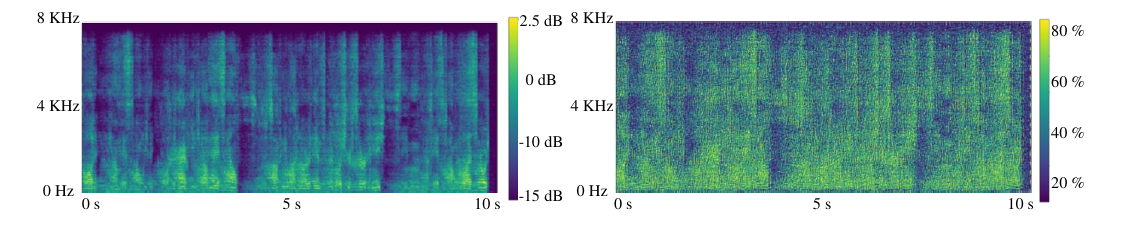}}
    \caption{Example of spectrogram and attention map on a noiseless sample of the synthetic dataset with a speaker talking at $10$ meters.}
    \label{fig:spec-heat}
\end{figure*}

\begin{figure*}[ht]
    \centering
\centerline{\includegraphics[width=1.10\linewidth]{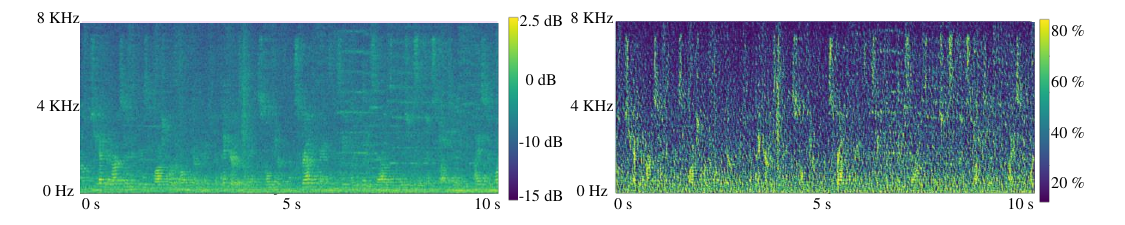}}
    \caption{Example of spectrogram and attention map on a noisy sample ($\mathrm{SNR} = 0$ dB) of the synthetic dataset with a speaker talking at $10$ meters.}
    \label{fig:spec-heat-0db}
\end{figure*}

Examples of noiseless and noisy spectrograms and attention maps are depicted in Figure~\ref{fig:spec-heat} and in Figure~\ref{fig:spec-heat-0db}, respectively. It is worth highlighting how the attention module differently focuses on the parts of the signal where the speech is most likely to stand out from the noise, or where the characteristics of the speech are still recognizable. In fact, the attention map in a noiseless case is evenly distributed across the entire frequency range since there is no noise that interferes.

\subsection{Convolutional Layers}
The architecture employs three convolutional blocks for feature extraction. In more detail, the structure of each block involves a 2D convolutional layer comprising $P_i$ $1\times 3$ filters, i.e., along the frequency axis with values of $8$, $32$, and $128$ assigned to the respective layers. We denote these filters as \textit{frequency kernels} whereas $3 \times 1$ filters are named \textit{time kernels}. Square kernels, known for their capability to capture time-frequency patterns, are commonly used in convolutional layers applied to spectrograms due to their effectiveness in capturing local patterns and structures along the frequency axis. In this work, the proposed model adopts rectangular filters, and temporal information is modeled by recurrent layers at the end of the model. In fact, rectangular filters can be more parameter-efficient compared to square kernels. Since the former has fewer parameters than square kernels of the same receptive field size, they can lead to a more compact model, making training and inference more computationally efficient and potentially reducing the risk of overfitting, especially when working with limited data.

Following this layer, a batch normalization~\cite{Ioffe_2015_ICML} step is applied, along with max and average pooling operations along the frequency dimension. Then, the results of which are summed.

The activation function utilized after each convolutional layer is the \ac{ELU}~\cite{Djork_2016_ICLR}, which is denoted as \begin{equation}
    \mathrm{ELU}(x) = \begin{cases}
  x, & x \geq 0 \\
  \alpha (e^x - 1) , & x < 0
\end{cases}
\end{equation} 
where $\alpha$ is a coefficient that regularizes the saturation of negative values. Notably, each layer employs a specific pooling rate denoted by $MP_i$, with values of $8$, $8$, and $2$ assigned to the respective layers. 

\subsection{Recurrent Layers}
To process the feature maps from the convolutional layers, two bi-directional \ac{GRU} layers are utilized with $\tanh(\cdot)$ as the activation function. These layers have exhibited promising results in audio and speech processing tasks, demonstrating parameter efficiency compared to \ac{LSTM} networks~\cite{Ravanelli_2018_TETCI}.

The output of the \ac{CNN} with shape $T \times 2 \times P$ is stacked along the channel dimension to produce a $T \times Q$ matrix to be fed to the recurrent layers. Then, in the proposed configuration, the extraction of reverberation-related information primarily relies on integrating information over time with the recurrent layers. Within this implementation, two bi-directional \acp{GRU} with $Q = 2P = 128$ neurons each for every time frame are employed.

Then, to predict the distance, three fully connected layers are employed, where an independent mapping between each time frame is performed in each layer. Firstly, the initial linear layer projects time-wise features from the last \ac{GRU} onto a matrix of dimensions $T\times R$, where $R = 128$. Subsequently, the second linear layer independently maps each time frame of the $T\times R$ matrix onto a vector of size $T \times 1$, denoted as the time-wise distance estimation $\mathbf{\hat{y}}$. Specifically, this vector represents the distance estimation for each time frame. Finally, the last fully connected layer is employed to perform regression and thus estimate the predicted distance, denoted as $\hat{y} \in \mathbb{R}$.

\subsection{Loss function}
The \ac{MSE} loss is used to train the \ac{DNN} system. Let $y \in \mathbb{R}$ be the true distance of a static sound source. In addition, let $\mathbf{y} \in \mathbb{R}^{T\times 1}$ be the vector consisting of frame-wise ground truth distances. Then, the loss used in the training phase for a single sample is
\begin{equation}\label{eq:loss}
    \mathcal{L}(y, \hat{y}, \mathbf{y}_t, \mathbf{\hat{y}}_t) = (y-\hat{y})^2 + ||\mathbf{y}_t-\mathbf{\hat{y}}_t||^2,
\end{equation}
where the loss is averaged across the batch dimension to be exploited by the backpropagation algorithm. Thanks to the imposition of the loss, the model predicts a distance for each time bin and, from this information, a single-valued distance. 
Having two losses in a static source scenario operates as a regularization term since it forces the proposed approach to return coherently both time-wise and single-distance estimations. However, in the context of dynamic sound sources, it is important to highlight that only frame-wise loss is required.

\begin{figure*}[htb]
\centering
\begin{tabular}{cc}
  \includegraphics[width=.485\textwidth]{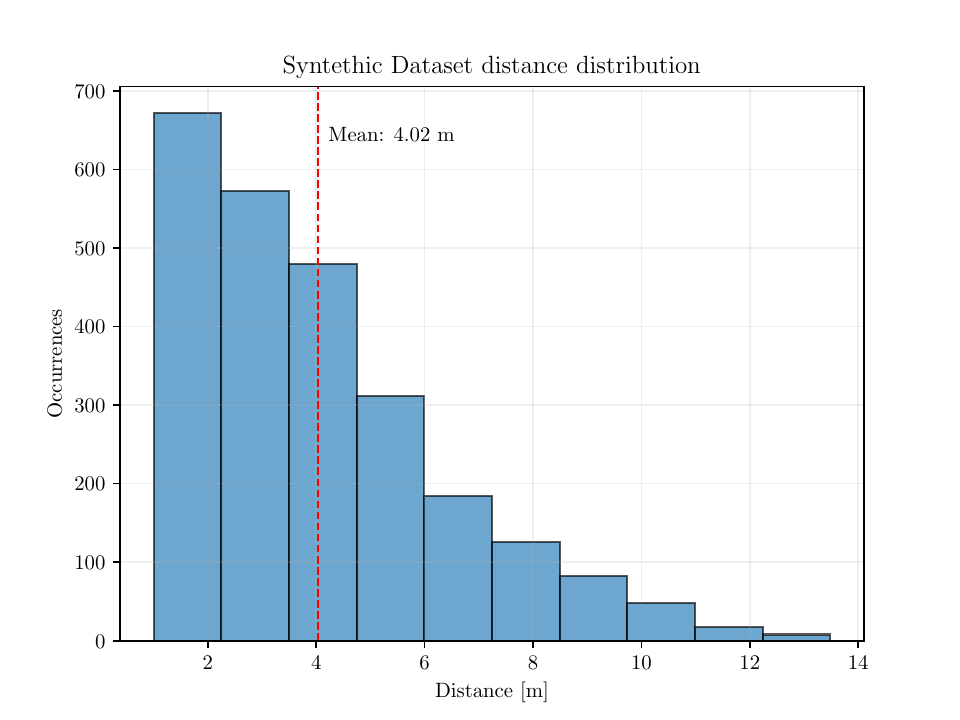}
  \includegraphics[width=.485\textwidth]{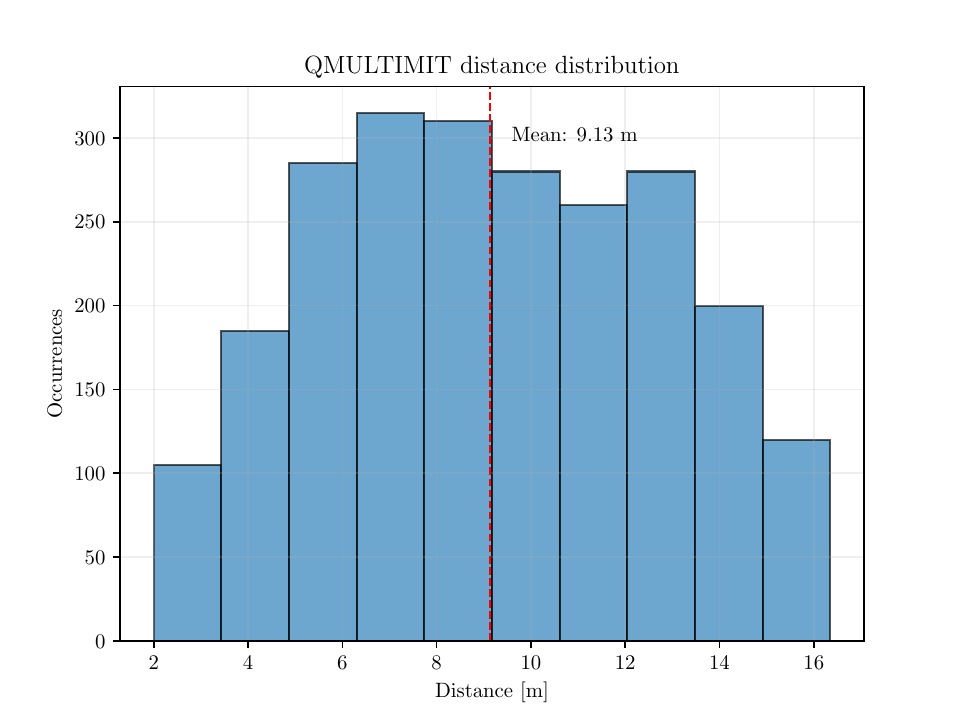} \\
  \includegraphics[width=.485\textwidth]{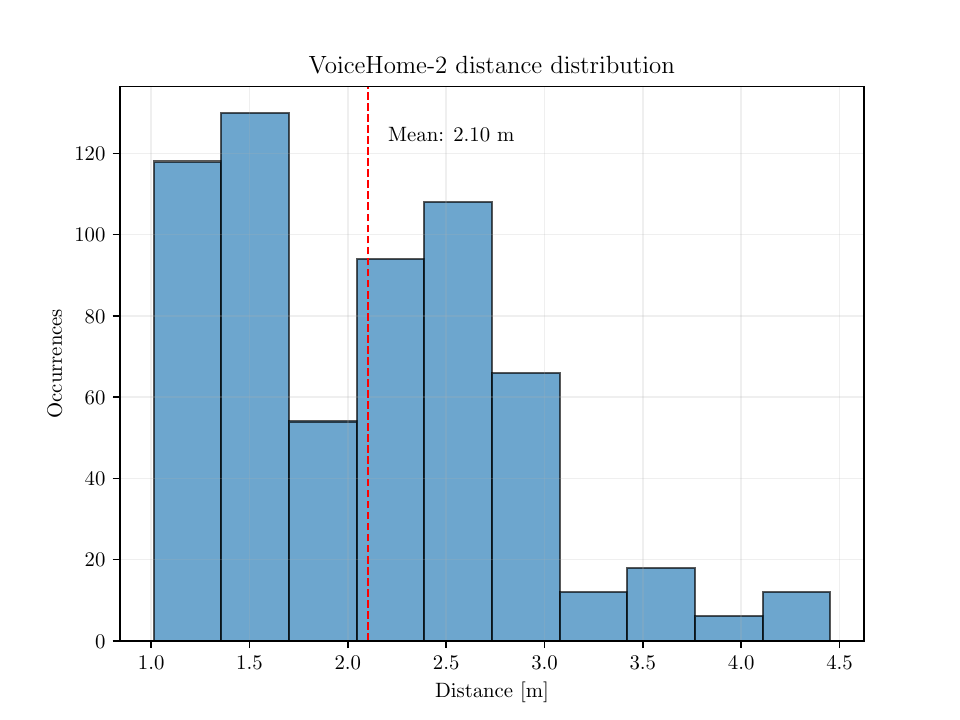} 
  \includegraphics[width=.485\textwidth]{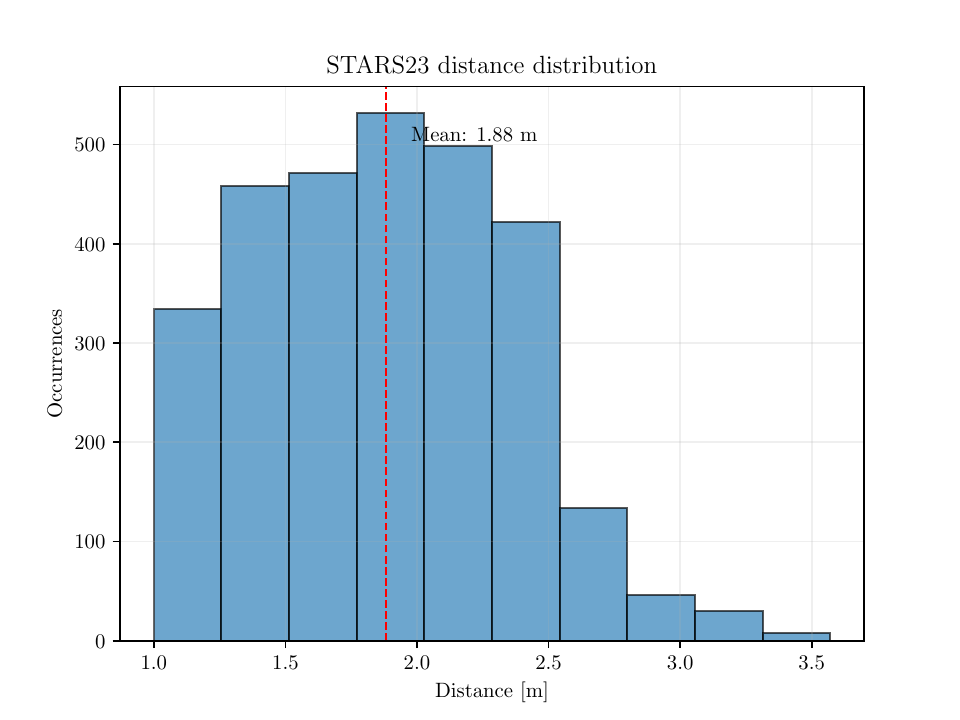}
    \end{tabular}
    \caption{Distributions of distances in each dataset.} 
    \label{fig:dist}
\end{figure*}

\subsection{Metrics}
The performance evaluation of our approach utilizes the \ac{MAE} ($\mathcal{L}_1$) as the performance measure for the entire test dataset

\begin{equation}
    \mathcal{L}_1 (y, \hat{y}) = |y-\hat{y}|,
\end{equation}
where the ground truth $y \in \mathbb{R}$ and the prediction $\hat{y} \in \mathbb{R}$ are considered. Additionally, the performance is assessed by calculating the \ac{MAE} within different distance ranges. This analysis allows us to quantify the relative error of our model concerning source distance. We define the relative \ac{MAE} ($ r \mathcal{L}_1$), which includes the real speaker distance in the evaluation, as follows:

\begin{equation}
    \mathrm{r}\mathcal{L}_1 (y, \hat{y}) = \frac{\mathcal{L}_1(y, \hat{y})}{y} =  \frac{|y-\hat{y}|}{y}.
\end{equation}
For the sake of clarity and brevity, \ac{MSE} has not been considered in the performance evaluation.

\section{Performance Assessment}\label{sec:perf}
This section describes how the performance assessment of the proposed approach has been carried out. To validate the work, three levels of realism have been addressed in the scope of speaker distance estimation:
\begin{itemize}
    \item \textbf{Synthetic}: simulated \acp{RIR} of an image-source room simulator are convolved with anechoic speech;
    \item \textbf{Hybrid}: measured \acp{RIR} are convolved with anechoic speech;
    \item \textbf{Real}: on-field reverberant speech recordings.
\end{itemize}
Figure~\ref{fig:dist} depicts the histograms of distances in each dataset employed in the experimental results.

\subsection{Synthetic Dataset}
The dataset used for experiments follows the same setup as in~\cite{Garcia_2022_EUSIPCO}. Briefly, anechoic speech recordings obtained from the TIMIT dataset~\cite{Garofalo_TIMIT_1993} are convolved with the simulated omnidirectional \acp{RIR} from an image-source room simulator for shoebox geometries~\cite{Politis_2016_PhD}. 

This simulator allows for frequency-dependent wall absorption and directional encoding of image sources in $5^{\mathrm{th}}$ order Ambisonics format. The elevation range between the source and the receiver spanned from $-35$° to $35$°. To compile a list of materials and their respective absorption coefficients for each surface type (ceiling, floor, and wall), we refer to widely used acoustical engineering tables~\cite{Tables_assorbing}. For each unique simulated room with its room-source-distance configuration, a random material is assigned to each surface, resulting in $2912$ possible material combinations. Compared to randomizing directly the target RT60 for each simulated room, this randomization approach allows us to avoid matching unnatural reverberation times to specific room volumes (e.g., a very long RT60 for a small room) and ensure a more natural distribution of reverberation times.

The final distribution of reverberation times exhibits a median, $10^{\mathrm{th}}$ percentile, and $90^{\mathrm{th}}$ percentile of $0.83$ s, $0.42$ s, and $2.38$ s, respectively. Furthermore, the positions of the sound sources are uniformly distributed in terms of the azimuth angle relative to the receiver.

The experiments include $2500$ audio files of $10$ s duration at $16$ kHz in compliance with the speech dataset. In the evaluation, $5$-fold cross validation is used where $1500$, $500$, and $500$ files are assigned to training, validation, and
testing in each fold.

To assess the performance of the proposed approach under different noise levels, real background noise is added into the synthetic dataset. Specifically, environmental noise recordings from the WHAM!~\cite{Wichern_2019_Interspeech} dataset, captured in various urban settings such as restaurants, cafes, and bars, are employed. Random segments of the same length as the simulated speech recordings are injected, mirroring the same split as the WHAM! dataset, with several \acp{SNR} levels ([$50$, $40$, $30$, $20$, $10$, $5$, $0$] dB). 


In addition to estimating the mean absolute distance estimation error, the errors are calculated separately for separate distance intervals that are $\{[1,2), [2,4), [4,8), [8, 14)\}$ meters. The \ac{MAE} errors are averaged using a $5$-fold cross-validation split, and the $95\%$ mean confidence intervals are evaluated.

\subsection{Hybrid Dataset - QMULTIMIT}
The \acp{RIR} used in the hybrid dataset, contained in the C4DM RIR database~\cite{Stewart_2010_ICASSP}, were measured in three rooms located at Queen Mary, University of London, London, UK. A Genelec 8250A loudspeaker was employed as the source for measuring all IRs, while each receiver position was measured using both an omnidirectional DPA 4006 and a B-format Soundfield SPS422B.

A collection of $130$ \acp{RIR} was captured in a classroom with dimensions $7.5 \times 9 \times 3.5$ m ($236$ $\mathrm{m}^3$) and consist of reflective surfaces such as a linoleum floor, painted plaster walls, ceiling, and a sizable whiteboard.

The second room, denoted as the Octagon, is a Victorian structure that was finalized in 1888. Presently serving as a conference venue, the walls of this building still showcase book-lined interiors, complemented by a wooden floor and plaster ceiling. As the name implies, this room features eight walls, each measuring $7.5$ m in length, and a domed ceiling towering $21$ m above the floor, resulting in an estimated volume of $9500$ $\mathrm{m}^3$. In the center of the room, a total of $169$ \acp{RIR} were measured.


The third room is The Great Hall which possesses a seating capacity of approximately $800$. It encompasses a stage and seating sections both on the floor and a balcony. To capture the audio, the microphones were positioned within the cleared seating area on the floor, spanning an area of approximately $23 \times 16$ m. The microphone placements mirror the layout used for the Octagon, encompassing $169$ \acp{RIR} over a $12 \times 12$ m region.

Following the same setup of the synthetic dataset, anechoic speech recordings are convolved from TIMIT~\cite{Garofalo_TIMIT_1993} and real background noises from WHAM!~\cite{Wichern_2019_Interspeech} are added with the measured \acp{RIR}, generating the hybrid QMULTIMIT dataset. For each \ac{RIR}, $5$ random speech recordings are selected from the TIMIT dataset, yielding $2340$ audio files.
\acp{RIR} are randomly divided into training, validation, and testing splits following a percentage ratio of $70$-$10$-$20$. Finally, the \ac{MAE} errors averaged across all the distance bins are provided. 

\subsection{Real Dataset}

\begin{description}[style=unboxed,leftmargin=0cm]
\item[\textbf{VoiceHome~-~2}~\cite{Bertin_2019_SpeechCommunication}.] This dataset is specifically made for distant speech processing applications in domestic environments. It consists of short commands for smart home devices in French, collected in reverberant conditions and uttered by twelve native French speakers facing the microphone.  The data is recorded in twelve different rooms corresponding to four houses, with fully annotated geometry, under quiet or noisy conditions. More precisely, VoiceHome~-~2 includes everyday noise sources (with no annotations regarding their \acp{SNR}) such as competing talkers, TV/radio, footsteps, doors, kitchenware, and electrical appliances.  Five speaker positions per room, comprising standing and sitting postures, are selected to encompass a broad range of angles and distances concerning the microphone array, which maintains a single, fixed position throughout all the room recordings. The sound is then captured by a microphone array consisting of eight \ac{MEMS} placed near the corner of a cubic baffle. For this study, only the first channel has been extracted. 
In total, VoiceHome~-~2 encompasses $752$ audio recordings, lasting approximately $10$ seconds for all the twelve rooms and the five noise scenes. 
The dataset is then randomly split using a percentage ratio of $70$-$10$-$20$ training, validation, and testing splits, respectively, for the experiments.

\item[\textbf{STARSS22}~\cite{Politis2022_dcase}.]  The dataset includes recordings of human interaction scenes with spatio-temporal event annotations for thirteen target classes, primarily focusing on speech. It is part of the DCASE Challenge 2022 Task 3 development set.  The recordings were made at two sites, Tampere University in Finland and Sony headquarters in Japan, in a total of eleven rooms maintaining a consistent organization and procedure regarding equipment, recording, and annotations. The dataset utilizes the Eigenmike spherical microphone array, offering two spatial formats. One format involves a tetrahedral sub-array of omnidirectional microphones mounted on a rigid spherical baffle. The corpus is more challenging compared to the other datasets due to the natural movement and orientation of multiple speakers during discussions, as well as the presence of intentional and unintentional sound events other than speech. It also contains diffuse and directional ambient noise at significant levels. Finally, audio data from a single microphone of the Eigenmike array has been processed, extracting $2934$ two-second single-speech excerpts that do not overlap with other annotated directional sources. 
As done before with the other datasets, STARSS22 is split using a percentage ratio of $70$-$10$-$20$ training, validation, and testing splits, respectively. 
\end{description}

It is worth noticing that, as can be inspected in Figure~\ref{fig:dist}, real dataset distances are differently distributed with respect to the synthetic and hybrid ones. The motivations of this behavior are as follows:
\begin{itemize}
    \item in many real-world scenarios, as in STARSS23~\cite{Politis_2023_ARXIV}, sound sources are not always at a fixed distance from the recording device;
    \item different recording environments can introduce variations in the speaker distance distribution. For example, in a controlled studio setting, speakers may be positioned at specific distances from the microphone to achieve desired sound characteristics. In contrast, field recordings or recordings made in everyday settings can have a wider range of distances due to the uncontrollable nature of the environment. Indeed, in this context, VoiceHome-2~\cite{Bertin_2019_SpeechCommunication} has been recorded in a domestic environment whereas STARSS23~\cite{Politis2022_dcase} has been collected in office-like environments;
    \item audio datasets are often curated to suit specific applications or scenarios. For instance, a dataset focused on speaker recognition in far-field scenarios may deliberately include more examples with distant speakers to simulate real-world challenges. On the other hand, a dataset for speech enhancement in close-proximity situations may prioritize examples with close speaker distances. VoiceHome~-~2 has been curately designed for enhancing distant-microphone speech whereas STARSS23 focuses on \ac{SELD}, yielding dissimilar distance distributions.
\end{itemize}

Accordingly with the distributions of distances in real scenarios, the distance bins used are $\{[1,2), [2,3), [3, 4.5)\}$ and $\{[1,2), [2,2.5), [2.5, 3)\}$ meters for VoiceHome~-~2 and STARSS22, respectively. The final \ac{MAE} errors are averaged using a percentage ratio of $70$-$10$-$20$ training, validation, and testing splits, respectively.

\section{Experimental Results}\label{sec:results}
In this section, the experimental results are shown for each realistic scenario, as detailed in Section~\ref{sec:perf}. First, the proposed architecture is tested on the synthetic dataset, both in noiseless and noisy scenarios, for the selection of hyperparameters. Next, the performance of the approach is evaluated on hybrid and real recordings by comparing the selected solution with different hyperparameters. Finally, an ablation study is provided to demonstrate the effectiveness of the attention module in all scenarios. 

\subsection{Implementation details}
For both training and fine-tuning procedures on all scenarios, the model is trained for $60$ epochs at a learning rate of $0.001$ with batch size of $16$ samples. A scheduled reduction ($80\%)$ of the learning rate is performed every $5$ epochs when the \ac{MSE} of the validation set does not improve. In this work, fine-tuning is carried out by training again the model, hence without the random initialization of the weights.

\begin{table*}[ht!]
\caption{Hyperparameters selection on the synthetic dataset with clean speech. The gray row highlights the proposed approach.}
\label{tab:syntheticClear}
    \centering
        \adjustbox{max width=1.02\textwidth}{%
    \begin{tabular}{lcc|cc|cc|cc|cc|cc}
    \hline \hline
    \multirow{2}*{Kernels} & \multirow{2}*{\# params} & \multirow{2}*{\# GRUs} & \multicolumn{2}{c|}{Average} & \multicolumn{2}{c|}{$[1,2)$} & \multicolumn{2}{c|}{$[2,4)$} & \multicolumn{2}{c|}{$[4, 8)$} & \multicolumn{2}{c}{$[8,14)$} \\
    
    & & & $\mathcal{L}_1$ & $\mathrm{r} \mathcal{L}_1$ & $\mathcal{L}_1$ & $\mathrm{r} \mathcal{L}_1$ & $\mathcal{L}_1$ & $\mathrm{r} \mathcal{L}_1$ & $\mathcal{L}_1$ & $\mathrm{r} \mathcal{L}_1$ & $\mathcal{L}_1$ & $\mathrm{r} \mathcal{L}_1$ \\
    \hline
    Binaural~\cite{Krause_TASLP_2024} & $650$ k & $2$ & $0.86 \pm 0.10$ & $0.29 \pm 0.05$ & $1.06 \pm 0.35$ & $0.72 \pm 0.22$ & $0.70 \pm 0.13$ & $0.25 \pm 0.05$ & $0.81 \pm 0.10$ & $0.15 \pm 0.02$ & $1.34 \pm 0.61$ & $0.13 \pm 0.05$ \\
    \hline \hline
    \textit{Time} & $123$ k & $0$ & $0.55 \pm 0.02$ & $0.18 \pm 0.01$ & $0.50 \pm 0.04$ & $0.35 \pm 0.03$ & $0.50 \pm 0.03$ & $0.18 \pm 0.01$ & $0.57 \pm 0.03$ & $0.11 \pm 0.01$ & $0.79 \pm 0.09$ & $0.08 \pm 0.01$ \\
    \textit{Squared} & $149$ k & $0$ & $0.70 \pm 0.02$ & $0.23 \pm 0.01$ & $0.59 \pm 0.04$ & $0.42 \pm 0.03$ & $0.68 \pm 0.04$ & $0.24 \pm 0.01$ & $0.71 \pm 0.04$ & $0.13 \pm 0.01$ & $1.03 \pm 0.12$ & $0.11 \pm 0.01$ \\
    \textit{Frequency} & $123$ k & $0$ & $0.86 \pm 0.03$ & $0.30 \pm 0.01$ & $0.83 \pm 0.05$ & $0.60 \pm 0.04$ & $0.80 \pm 0.04$ & $0.28 \pm 0.02$ & $0.86 \pm 0.04$ & $0.16 \pm 0.01$ & $1.17 \pm 0.14$ & $0.12 \pm 0.01$ \\
    \hline \hline
    \textit{Time} & $353$ k  & $1$ & $0.16 \pm 0.01$ & $0.05 \pm 0.00$ & $0.15 \pm 0.01$ & $0.11 \pm 0.01$ & $0.13 \pm 0.01$ & $0.05 \pm 0.00$ & $0.19 \pm 0.01$ & $0.03 \pm 0.00$ & $0.27 \pm 0.03$ & $0.03 \pm 0.00$  \\
    \textit{Squared} & $379$ k  & $1$ & $0.15 \pm 0.01$ & $0.05 \pm 0.00$ & $0.13 \pm 0.01$ & $0.09 \pm 0.01$ & $0.11 \pm 0.01$ & $0.04 \pm 0.00$ & $0.16 \pm 0.01$ & $0.03 \pm 0.00$ & $0.27 \pm 0.04$ & $0.03 \pm 0.00$ \\
    \textit{Frequency} & $353$ k & $1$ & $0.13 \pm 0.01$ & $\mathbf{0.04 \pm 0.00}$ & $\mathbf{0.12 \pm 0.01}$ & $\mathbf{0.08 \pm 0.01}$ & $0.10 \pm 0.01$ & $0.04 \pm 0.00$ & $0.13 \pm 0.01$ & $\mathbf{0.02 \pm 0.00}$ & $0.24 \pm 0.04$ & $\mathbf{0.02 \pm 0.00}$  \\
    \hline \hline
    
    \textit{Time} & $650$ k & $2$ & $0.13 \pm 0.01$ & $\mathbf{0.04 \pm 0.00}$ & $\mathbf{0.12 \pm 0.01}$ & $0.09 \pm 0.01$ & $0.10 \pm 0.01$ & $0.04 \pm 0.00$ & $0.13 \pm 0.01$ & $\mathbf{0.02 \pm 0.00}$ & $0.24 \pm 0.07$ & $\mathbf{0.02 \pm 0.01}$ \\
    
    \textit{Squared} & $676$ k & $2$ & $\mathbf{0.11 \pm 0.00}$ & $\mathbf{0.04 \pm 0.00}$ & $\mathbf{0.12 \pm 0.01}$ & $\mathbf{0.08 \pm 0.01}$ & $\mathbf{0.09 \pm 0.00}$ & $\mathbf{0.03 \pm 0.00}$ & $0.12 \pm 0.01$ & $\mathbf{0.02 \pm 0.00}$ & $0.18 \pm 0.03$ & $\mathbf{0.02 \pm 0.00}$ \\

    \rowcolor{Gray} \textit{Frequency} & $650$ k & $2$ & $\mathbf{0.11 \pm 0.00}$ & $\mathbf{0.04 \pm 0.00}$ & $\mathbf{0.12 \pm 0.01}$ & $\mathbf{0.08 \pm 0.01}$ & $0.10 \pm 0.00$ & $\mathbf{0.03 \pm 0.00}$ & $\mathbf{0.11 \pm 0.01}$ & $\mathbf{0.02 \pm 0.00}$ & $\mathbf{0.16 \pm 0.02}$ & $\mathbf{0.02 \pm 0.00}$ \\
     \hline \hline
    \end{tabular}
}
\end{table*}

\subsection{Results on noiseless synthetic data}

The proposed approach efficiently estimates speaker distance with an average error of $11$ cm in a noiseless scenario, as it can be inspected from Table~\ref{tab:syntheticClear}. Since there is no other published method that attempts regression-based SDE with a single microphone, for comparison purposes we present results on binaural SDE following the recently published work of ~\cite{Krause_TASLP_2024}. The binaural estimation model is similar to the CRNN model used herein; however, we modify it to include the attention operation proposed in this work for better comparison purposes. A similar simulator, range of acoustic conditions, and number of rooms was used in~\cite{Krause_TASLP_2024} as herein. The same spectrogram and binaural features are also used as in the original work. The binaural estimation results ($86$ cm) we obtain are, on average, better than the ones in~\cite{Krause_TASLP_2024} ($151$ cm), with the improvement most likely attributed to the use of the attention layers. However, the most striking difference is that of the monophonic omnidirectional results ($11$ cm) versus the binaural ones ($86$ cm). It seems that the complex frequency-, direction-, and orientation-dependent effects imposed by head-related transfer functions (HRTFs) make it harder for the model to associate spectrotemporal reverberation patterns with the source distance. However, a definite conclusion on differences between single-channel omnidirectional versus binaural SDE requires further study.

An increasing trend of the errors with respect to the distance is notable. This behavior is expected due to the dominant influence of the late reverberant component compared to the direct and early reflection components of the signal at long distances. These late reverberation cues exhibit statistical diffusion~\cite{jacobsen2000coherence}, meaning that short-term magnitudes and phases resemble noise-like characteristics. Consequently, extracting meaningful information from these dominant late reverberation cues may pose challenges for the model in effectively estimating speaker distance.

\begin{figure}[ht!]
    \centering
    \includegraphics[width=1\linewidth]{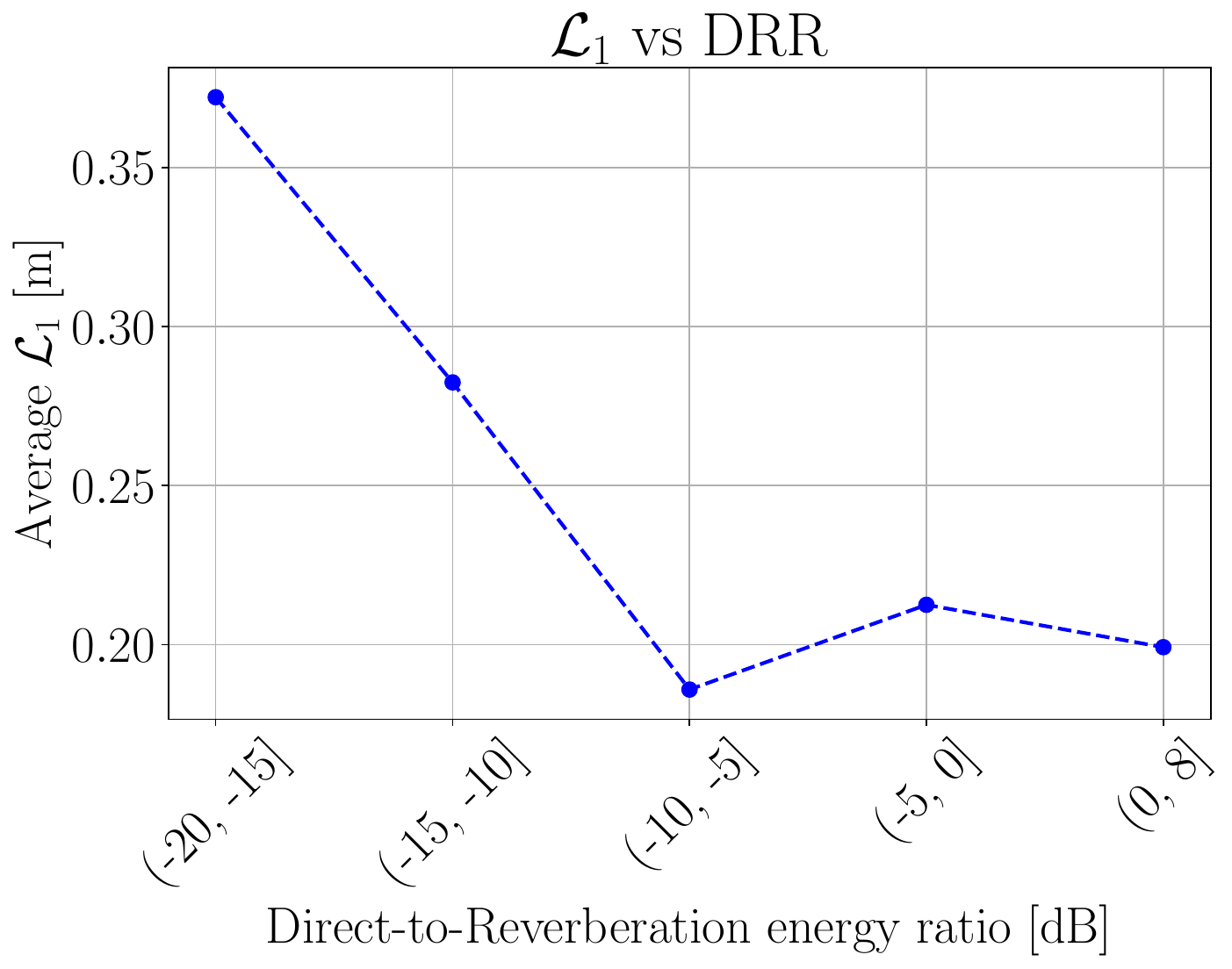}
    \caption{Relation between \ac{DRR} and $\mathcal{L}_1$.}
    \label{fig:L1_DRR}
\end{figure}

Such behaviour is demonstrated in Figure~\ref{fig:L1_DRR}. Considering that the balance between direct speech energy versus early and late reverberant energy is exemplified in the \ac{DRR}, measured from the simulated RIRs, it is clear that dominance of the reverberation at low DRRs impacts negatively distance estimation. There seems to be an optimum balance where both direct sound and reverberation contribute to estimation, after which direct sound can start to mask reverberation-related cues for higher DRRs, with a subsequent small drop in performance. A closer investigation of distance estimation at very high DRRs or very small distances at the near-field of the microphone is left for future work.

Moreover, the results of the study demonstrate that the \ac{GRU} layers play a crucial role in the model's performance. The \ac{GRU} layers likely contribute to the model's ability to capture sequential patterns and dependencies effectively. Additionally, the study found that using rectangular kernels, as opposed to square kernels, in combination with \ac{GRU} layers improves the model's efficiency. In this scenario, the rectangular kernels are better at capturing different types of patterns and features in the data, leading to more effective and efficient information processing within the model. This statement, however, does not hold when no \ac{GRU} layers are present. 

In addition, it is worth noting that using a single \ac{GRU} layer slightly impacts the overall performance of the proposed approach, approximately halving the number of learnable parameters.

\begin{table}[ht!]
\caption{Experimental results on noisy synthetic data with fixed \ac{SNR} and frequency kernels. Gray row highlights the proposed approach.}
\label{tab:impactPhaseFeatures}
    \centering
        \adjustbox{max width=\textwidth}{%
    \begin{tabular}{cc|cc}
    \hline \hline
     \ac{SNR} & Feature set & $\mathcal{L}_1$ & $\mathrm{r} \mathcal{L}_1$ \\
    \hline
    \multirow{3}*{$50$ dB} & w\slash $|\mathrm{STFT}|$ & $0.48 \pm 0.02$ & $0.14 \pm 0.01$  \\
    & w\slash sinus and cosinus & $\mathbf{0.37 \pm 0.02}$ & $\mathbf{0.11 \pm 0.01}$   \\
    \rowcolor{Gray} & $|\mathrm{STFT}|$ + sinus and cosinus & $0.41 \pm 0.02$ & $0.12 \pm 0.00$ \\
    \hline \hline 
    \multirow{3}*{$40$ dB} & w\slash $|\mathrm{STFT}|$ & $0.77 \pm 0.03$ & $\mathbf{0.21 \pm 0.01}$ \\
    & w\slash sinus and cosinus & $\mathbf{0.71 \pm 0.03}$ & $\mathbf{0.21 \pm 0.01}$ \\
    \rowcolor{Gray} & $|\mathrm{STFT}|$ + sinus and cosinus  & $0.87 \pm 0.04$ & $0.24 \pm 0.01$ \\
    \hline \hline 
    \multirow{3}*{$30$ dB} & w\slash $|\mathrm{STFT}|$ & $\mathbf{1.11 \pm 0.04}$ & $\mathbf{0.30 \pm 0.01}$ \\
    & w\slash sinus and cosinus & $1.51 \pm 0.06$ & $0.45 \pm 0.02$  \\
    \rowcolor{Gray} & $|\mathrm{STFT}|$ + sinus and cosinus & $1.14 \pm 0.04$ & $0.31 \pm 0.01$ \\
    \hline \hline 
    \multirow{3}*{$20$ dB} & w\slash $|\mathrm{STFT}|$ & $\mathbf{1.20 \pm 0.04}$ & $\mathbf{0.33 \pm 0.01}$ \\
    & w\slash sinus and cosinus & $1.76 \pm 0.06$ & $0.56 \pm 0.02$  \\
    \rowcolor{Gray} & $|\mathrm{STFT}|$ + sinus and cosinus  & $1.21 \pm 0.05$ & $\mathbf{0.33 \pm 0.01}$  \\
    \hline \hline 
    \multirow{3}*{$10$ dB} & w\slash $|\mathrm{STFT}|$ & $1.30 \pm 0.05$ & $0.36 \pm 0.01$ \\
    & w\slash sinus and cosinus & $1.70 \pm 0.06$ & $0.56 \pm 0.02$  \\
    \rowcolor{Gray} & $|\mathrm{STFT}|$ + sinus and cosinus  & $\mathbf{1.27 \pm 0.05}$ & $\mathbf{0.35 \pm 0.01}$ \\
    \hline \hline 
    \multirow{3}*{$5$ dB} & w\slash $|\mathrm{STFT}|$ & $1.34 \pm 0.05$ & $0.38 \pm 0.01$  \\
    & w\slash sinus and cosinus & $1.73 \pm 0.06$ & $0.58 \pm 0.02$ \\
    \rowcolor{Gray} & $|\mathrm{STFT}|$ + sinus and cosinus  & $\mathbf{1.26 \pm 0.05}$ & $\mathbf{0.34 \pm 0.01}$  \\
    \hline \hline 
    \multirow{3}*{$0$ dB} & w\slash $|\mathrm{STFT}|$ & $1.47 \pm 0.05$ & $0.44 \pm 0.02$ \\
    & w\slash sinus and cosinus & $1.77 \pm 0.06$ & $0.61 \pm 0.02$ \\
    \rowcolor{Gray} & $|\mathrm{STFT}|$ + sinus and cosinus  & $\mathbf{1.39 \pm 0.05}$ & $\mathbf{0.42 \pm 0.02}$  \\
    \hline \hline
    \end{tabular}
}
\end{table}

\subsection{Analysis of the impact of noise on synthetic data}

\begin{figure}[ht]
    \centering
    \includegraphics[width=1\linewidth]{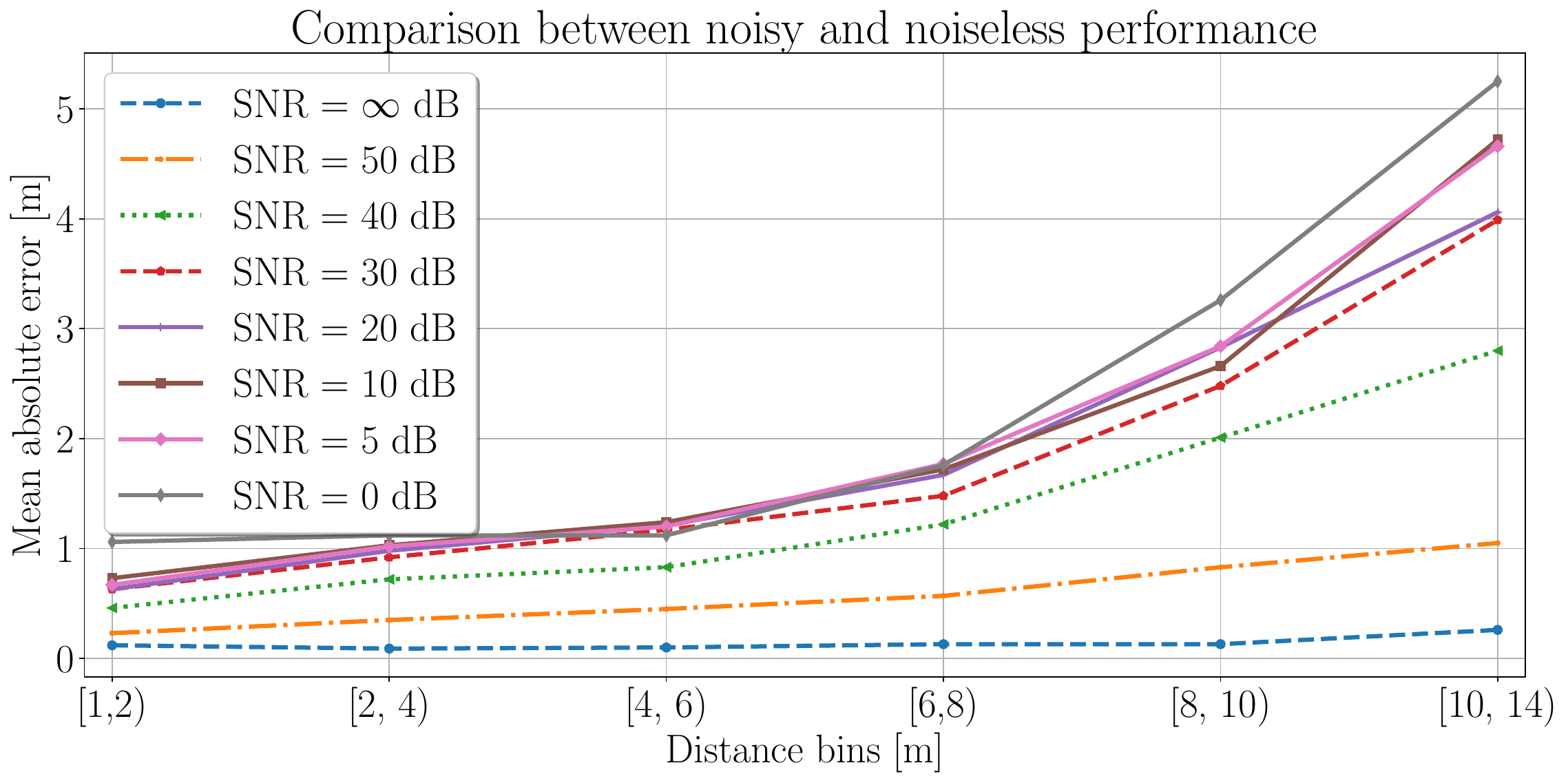}
    \caption{Comparison between noisy and noiseless performance of
the proposed approach on the synthetic dataset.}
    \label{fig:errorsSNR}
\end{figure}

To assess the quality of the predictions in relation to noise strength, seven \ac{SNR} values have been specifically chosen during training. More precisely, a separate model is trained from scratch for each \ac{SNR} level. Table~\ref{tab:impactPhaseFeatures} depicts the results where a notable discrepancy between the noiseless and noisy scenarios becomes evident. This divergence is primarily attributed to the disruptive influence of background noise on the phase information~\cite{Neri_2023_WASPAA}, which has been also demonstrated in speech enhancement studies~\cite{paliwal2011importance}. It is worth noting from Figure~\ref{fig:errorsSNR} that the performance of the proposed method remains consistent across all \ac{SNR} levels for distances up to $6$ meters.

However, beyond this distance, the error increases rapidly. This behavior can be attributed to the quadratic inverse relationship between distance and sound intensity, i.e., $I_s \propto \frac{1}{d^2}$. Due to this physical behavior, the direct sound and early distinct echoes exhibit similar energy levels compared to the late reverberant cues, hindering long-distance information.

\subsection{Results on hybrid data}
As done with the synthetic dataset, five \ac{SNR} values have been selected to assess the performance of the proposed architecture by training a separate model from scratch for each \ac{SNR} level. Table~\ref{tab:hybrid} shows the experimental results, highlighting the superiority of the chosen configuration. The notation $[30, +\infty)$ dB denotes the results of the model both in noiseless case and with at most $30$ dB of \ac{SNR}. It is worth noting that, differently from the synthetic scenario, the impact of background noise is smaller even at low \ac{SNR}. In fact, comparing Table~\ref{tab:impactPhaseFeatures} with Table~\ref{tab:hybrid}, it is evident how synthetic \acp{RIR} are more affected by noise at higher \ac{SNR} with respect to measured ones.

Interestingly, the use of only sinus and cosinus maps yields poor performance at all \acp{SNR} levels whereas the \ac{STFT} magnitude is essential for the task. This result agrees with the previous study~\cite{Neri_2023_WASPAA} where the use of only sinus and cosinus features in noisy audio recordings is ineffective.

\begin{table}[ht!]
\caption{Distance estimation errors for the QMULTIMIT hybrid dataset. Gray row highlights the proposed approach. All features are used if not mentioned}.
\label{tab:hybrid}
    \centering
        \adjustbox{max width=0.5\textwidth}{%
    \begin{tabular}{cl|ccc}
    \hline \hline
    SNR & Hyperparameters & \# GRUs & $\mathcal{L}_1$ & $\mathrm{r} \mathcal{L}_1$\\
    \hline \hline 
    \multirow{11}*{$[30,+ \infty)$ dB} &  \textit{Time} & 0 & $2.49 \pm 0.16$ & $0.28 \pm 0.02$ \\
    & \textit{Squared}  & 0 & $2.38 \pm 0.15$ & $0.25 \pm 0.02$ \\
    & \textit{Frequency}  & 0 & $2.97 \pm 0.17$ & $0.33 \pm 0.03$  \\
    \cline{2-5}
    & \textit{Time} & 1 & $1.58 \pm 0.12$ & $0.16 \pm 0.01$ \\
    & \textit{Squared}  & 1 & $1.52 \pm 0.12$ & $0.15 \pm 0.01$ \\
    & \textit{Frequency}  & 1 & $1.68 \pm 0.12$ & $0.17 \pm 0.01$  \\
    \cline{2-5}
    & \textit{Time} & 2  & $1.70 \pm 0.12$ & $0.17 \pm 0.01$ \\
    & \textit{Squared} & 2 & $\mathbf{1.48 \pm 0.13}$ & $\mathbf{0.14 \pm 0.01}$ \\
    & \textit{Freq.} w\slash $|\mathrm{STFT}|$  & 2 & $1.67 \pm 0.13$ & $0.17 \pm 0.01$ \\ 
    & \textit{Freq.} w\slash sinus and cosinus & 2 & $2.17 \pm 0.14$ & $0.23 \pm 0.02$\\
    \rowcolor{Gray} & \textit{Frequency} & 2 & $1.52 \pm 0.12$ & $0.15 \pm 0.01$  \\
    \hline \hline 
    \multirow{11}*{$20$ dB} & \textit{Time} & 0 & $2.22 \pm 0.15$ & $0.24 \pm 0.02$  \\
    & \textit{Squared} & 0 & $2.36 \pm 0.15$ & $0.25 \pm 0.02$ \\
    & \textit{Frequency}  & 0 & $2.88 \pm 0.17$ & $0.32 \pm 0.02$ \\
    \cline{2-5}
    & \textit{Time} & 1 & $1.67 \pm 0.12$ & $0.16 \pm 0.01$ \\
    & \textit{Squared}  & 1 & $\mathbf{1.46 \pm 0.12}$ & $\mathbf{0.14 \pm 0.01}$ \\
    & \textit{Frequency}  & 1 & $1.71 \pm 0.12$ & $0.17 \pm 0.01$  \\
    \cline{2-5}
    & \textit{Time}  & 2  & $1.66 \pm 0.13$ & $0.16 \pm 0.01$ \\
    & \textit{Squared}  & 2 & $1.60 \pm 0.13$ & $0.16 \pm 0.01$ \\
    & \textit{Freq.} w\slash $|\mathrm{STFT}|$ & 2 & $1.64 \pm 0.13$ & $0.16 \pm 0.01$ \\
    & \textit{Freq.} w\slash sinus and cosinus & 2 & $1.98 \pm 0.13$ & $0.21 \pm 0.02$ \\
    \rowcolor{Gray} & \textit{Frequency} & 2 & $1.48 \pm 0.11$ & $\mathbf{0.14 \pm 0.01}$ \\
    \hline \hline 
    \multirow{11}*{$10$ dB} & \textit{Time}  & 0 & $2.23 \pm 0.14$ & $0.24 \pm 0.02$ \\
    & \textit{Squared} & 0 & $2.20 \pm 0.14$ & $0.24 \pm 0.02$ \\
    & \textit{Frequency} & 0 & $2.55 \pm 0.14$ & $0.28 \pm 0.02$  \\
    \cline{2-5}
    & \textit{Time} & 1 & $1.71 \pm 0.12$ & $0.17 \pm 0.01$  \\
    & \textit{Squared}  & 1 & $1.58 \pm 0.13$ & $0.16 \pm 0.01$ \\
    & \textit{Frequency}  & 1 & $1.60 \pm 0.12$ & $0.16 \pm 0.01$ \\
    \cline{2-5}
    & \textit{Time} & 2 & $1.65 \pm 0.12$ & $0.16 \pm 0.01$ \\
    & \textit{Squared}  & 2 & $1.56 \pm 0.13$ & $\mathbf{0.15 \pm 0.01}$ \\
    & \textit{Freq.} w\slash $|\mathrm{STFT}|$ & 2 & $\mathbf{1.55 \pm 0.12}$ & $\mathbf{0.15 \pm 0.01}$ \\
    & \textit{Freq.} w\slash sinus and cosinus & 2 & $1.97 \pm 0.12$ & $0.21 \pm 0.01$ \\
    \rowcolor{Gray} & \textit{Frequency} & 2 & $1.65 \pm 0.13$ & $0.17 \pm 0.01$ \\
    \hline \hline 
    \multirow{11}*{$0$ dB} & \textit{Time}  & 0 & $2.54 \pm 0.14$ & $0.28 \pm 0.02$ \\
    & \textit{Squared}  & 0 & $2.74 \pm 0.15$ & $0.30 \pm 0.02$  \\
    & \textit{Frequency}   & 0 & $3.01 \pm 0.15$ & $0.33 \pm 0.02$  \\
    \cline{2-5}
    & \textit{Time} & 1 & $1.75 \pm 0.12$ & $0.18 \pm 0.01$ \\
    & \textit{Squared}  & 1 & $1.83 \pm 0.12$ & $0.19 \pm 0.01$ \\
    & \textit{Frequency}  & 1 & $1.82 \pm 0.13$ & $0.19 \pm 0.01$ \\
    \cline{2-5}
    & \textit{Time}  & 2 & $2.46 \pm 0.15$ & $0.23 \pm 0.01$   \\
    & \textit{Squared}   & 2 & $1.98 \pm 0.12$ & $0.21 \pm 0.02$  \\
    & \textit{Freq.} w\slash $|\mathrm{STFT}|$ & 2 & $\mathbf{1.63 \pm 0.13}$ & $\mathbf{0.17 \pm 0.01}$ \\
    & \textit{Freq.} w\slash sinus and cosinus & 2 & $2.24 \pm 0.13$ & $0.25 \pm 0.02$ \\
    \rowcolor{Gray} & \textit{Frequency} & 2 & $1.66 \pm 0.13$ & $\mathbf{0.17 \pm 0.01}$ \\
    \hline \hline 
    \multirow{11}*{$-10$ dB} & \textit{Time} & 0 & $3.03 \pm 0.14$ & $0.34 \pm 0.03$ \\
    & \textit{Squared} & 0 & $3.03 \pm 0.14$ & $0.33 \pm 0.02$  \\
    & \textit{Frequency}  & 0 & $3.04 \pm 0.14$ & $0.33 \pm 0.02$ \\
    \cline{2-5}
    & \textit{Time} & 1 & $3.02 \pm 0.14$ & $0.33 \pm 0.02$ \\
    & \textit{Squared}  & 1 & $3.01 \pm 0.14$ & $0.33 \pm 0.03$ \\
    & \textit{Frequency}  & 1 & $3.00 \pm 0.14$ & $0.33 \pm 0.03$ \\
    \cline{2-5}
    & \textit{Time}  & 2 & $3.06 \pm 0.14$ & $0.34 \pm 0.03$ \\
    & \textit{Squared}  & 2 & $2.57 \pm 0.13$ & $0.28 \pm 0.02$  \\
    & \textit{Freq.} w\slash $|\mathrm{STFT}|$ & 2 & $\mathbf{2.28 \pm 0.13}$ & $\mathbf{0.25 \pm 0.02}$  \\
    & \textit{Freq.} w\slash sinus and cosinus & 2 & $3.01 \pm 0.14$ & $0.33 \pm 0.03$\\
    \rowcolor{Gray} & \textit{Frequency} & 2 & $2.34 \pm 0.13$ & $\mathbf{0.25 \pm 0.02}$ \\
    \hline
    \hline 
    \end{tabular}
}
\end{table}

\subsection{Results on real data}
Table~\ref{tab:RealDataVoiceHome} and Table~\ref{tab:STARS} depict the results on VoiceHome~-~2~\cite{Bertin_2019_SpeechCommunication} and STARSS23~\cite{Politis2022_dcase}, respectively. Following the same rationale of the synthetic and hybrid scenarios, the selected configuration outperforms the other models. The results obtained from the analysis of real data demonstrate the clear superiority of the proposed model in accurately estimating distances. Across both datasets, the proposed model consistently outperforms different configurations of the models, showcasing its robustness and effectiveness. However, it is worth noting that a few outliers surfaced in the results, particularly within the VoiceHome~-~2 dataset where large confidence intervals are present. This occurrence can be attributed to the limited size of the datasets as the model overfits the training dataset. With a larger dataset, these outliers are expected to be mitigated, and the model's performance is likely to become even more reliable and precise. This observation underscores the potential for further advancement in distance estimation when working with more extensive datasets.

\subsection{Ablation study of the attention module}

To demonstrate the effectiveness of the attention module, an ablation study is performed on all the scenarios. First, performance assessment is carried out without the module. Then, instead of returning a $T \times F \times 3$ matrix, a spectrogram attention map, i.e., $T \times F$, is learned by a module. Then, an element-wise multiplication is performed between the magnitude of the \ac{STFT} and the attention map. 

These three modalities are analyzed in Table~\ref{tab:ablationAttention}, depicting the errors for each bin with their confidence intervals. Predicting an attention map for each feature provides better distance estimation on average.
Moreover, the results demonstrate that all the approaches perform similarly in the short range, up to $8$ meters. Conversely, applying the attention map on each of the feature maps in the feature set produces better outcomes in the long range with respect to the other two cases. When the speaker is far from the microphone, the learned attention maps enhance the features set, facilitating the extraction of features of the convolutional layers. Indeed, as the distance between the speaker and the microphone increases, detecting these patterns becomes more challenging due to their reduced salience~\cite{jacobsen2000coherence}.

Moreover, an ablation study has been carried out also on the hybrid and real data, as it can be inspected in Table~\ref{tab:ablationAttentionHybridReal}. The attention map yields the best performance in the hybrid case when it is only applied to the \ac{STFT} magnitude channel. This fact highlights the ineffectiveness of phase features in this specific use case. Instead, the results demonstrate the superiority of the attention map applied on all the channels in the real scenario.

\begin{table*}[ht!]
\caption{Distance estimation errors for the VoiceHome~-~2 dataset. Gray row highlights the proposed approach. All features are used if not mentioned}
\label{tab:RealDataVoiceHome}
    \centering
        \adjustbox{max width=\textwidth}{%
    \begin{tabular}{lc|cc|cc|cc|cc}
    \hline \hline
    \multirow{2}*{Hyperparameters} & \multirow{2}*{\# GRUs} & \multicolumn{2}{c|}{Average} & \multicolumn{2}{c|}{$[1,2)$} & \multicolumn{2}{c|}{$[2,3)$} & \multicolumn{2}{c}{$[3, 4.5)$} \\
    & & $\mathcal{L}_1$ & $\mathrm{r} \mathcal{L}_1$ & $\mathcal{L}_1$ & $\mathrm{r} \mathcal{L}_1$ & $\mathcal{L}_1$ & $\mathrm{r} \mathcal{L}_1$ & $\mathcal{L}_1$ & $\mathrm{r} \mathcal{L}_1$ \\
    
    \hline 
    \textit{Time} & 0 & $0.95 \pm 0.10$ & $0.49 \pm 0.06$ & $1.00 \pm 0.14$ & $0.73 \pm 0.11$ & $0.69 \pm 0.12$ & $0.28 \pm 0.05$ & $1.20 \pm 0.25$ & $0.32 \pm 0.06$ \\
    \textit{Squared} & 0 & $0.90 \pm 0.11$ & $0.46 \pm 0.07$ & $0.90 \pm 0.16$ & $0.69 \pm 0.14$ & $0.57 \pm 0.11$ & $0.23 \pm 0.04$ & $1.34 \pm 0.26$ & $0.35 \pm 0.07$ \\
    \textit{Frequency} & 0 & $0.83 \pm 0.09$ & $0.43 \pm 0.06$ & $0.85 \pm 0.13$ & $0.63 \pm 0.11$ & $0.67 \pm 0.13$ & $0.27 \pm 0.05$ & $1.02 \pm 0.20$ & $0.27 \pm 0.05$ \\
    \hline
    \textit{Time} & 1 & $0.76 \pm 0.09$ & $0.38 \pm 0.05$ & $0.73 \pm 0.12$ & $0.55 \pm 0.10$ & $0.47 \pm 0.10$ & $0.19 \pm 0.04$ & $1.19 \pm 0.23$ & $0.32 \pm 0.06$ \\
    \textit{Squared} & 1 & $0.74 \pm 0.09$ & $0.40 \pm 0.07$ & $0.85 \pm 0.15$ & $0.65 \pm 0.13$ & $0.43 \pm 0.09$ & $\mathbf{0.17 \pm 0.04}$ & $0.96 \pm 0.20$ & $0.26 \pm 0.05$ \\
    \textit{Frequency} & 1 & $0.74 \pm 0.08$ & $0.37 \pm 0.05$ & $0.73 \pm 0.12$ & $0.54 \pm 0.10$ & $0.53 \pm 0.10$ & $0.21 \pm 0.04$ & $1.06 \pm 0.21$ & $0.28 \pm 0.05$ \\
    \hline
    \textit{Time} & 2 & $0.64 \pm 0.08$ & $\mathbf{0.31 \pm 0.05}$ & $\mathbf{0.59 \pm 0.12}$ & $\mathbf{0.44 \pm 0.10}$ & $0.49 \pm 0.09$ & $0.20 \pm 0.03$ & $0.94 \pm 0.21$ & $0.25 \pm 0.05$ \\
    \textit{Squared} & 2 & $0.70 \pm 0.10$ & $0.35 \pm 0.06$ & $0.67 \pm 0.14$ & $0.51 \pm 0.12$ & $\mathbf{0.43 \pm 0.12}$ & $\mathbf{0.17 \pm 0.05}$ & $1.11 \pm 0.21$ & $0.29 \pm 0.05$ \\
    \textit{Freq} w \slash $|\mathrm{STFT}|$  & 2 & $0.66 \pm 0.08$ & $0.33 \pm 0.05$ & $0.63 \pm 0.13$ & $0.48 \pm 0.11$ & $0.47 \pm 0.10$ & $0.19 \pm 0.04$ & $0.98 \pm 0.17$ & $0.27 \pm 0.05$ \\
    \textit{Freq} w \slash sinus and cosinus  & 2 & $0.91 \pm 0.11$ & $0.46 \pm 0.07$ & $0.88 \pm 0.14$ & $0.68 \pm 0.13$ & $0.52 \pm 0.11$ & $0.21 \pm 0.04$ & $1.49 \pm 0.21$ & $0.40 \pm 0.05$ \\
    \rowcolor{Gray} \textit{Frequency} & 2 & $\mathbf{0.63 \pm 0.08}$ & $0.32 \pm 0.05$ & $0.64 \pm 0.11$ & $0.48 \pm 0.10$ & $0.48 \pm 0.11$ & $0.19 \pm 0.04$ & $\mathbf{0.80 \pm 0.20}$ & $\mathbf{0.21 \pm 0.05}$ \\
    \hline \hline
    \end{tabular}
}
\end{table*}

\begin{table*}[ht!]
\caption{Distance estimation errors for the STARSS23 dataset. Gray row highlights the proposed approach. All features are used if not mentioned}
\label{tab:STARS}
    \centering
        \adjustbox{max width=\textwidth}{%
    \begin{tabular}{lc|cc|cc|cc|cc}
    \hline \hline
    \multirow{2}*{Hyperparameters} & \multirow{2}*{\# GRUs} & \multicolumn{2}{c|}{Average} & \multicolumn{2}{c|}{$[1,2)$} & \multicolumn{2}{c|}{$[2,2.5)$} & \multicolumn{2}{c}{$[2.5, 3)$} \\
    & & $\mathcal{L}_1$ & $\mathrm{r} \mathcal{L}_1$ & $\mathcal{L}_1$ & $\mathrm{r} \mathcal{L}_1$ & $\mathcal{L}_1$ & $\mathrm{r} \mathcal{L}_1$ & $\mathcal{L}_1$ & $\mathrm{r} \mathcal{L}_1$ \\
    \hline
    \textit{Time} & 0 & $0.51 \pm 0.03$ & $0.23 \pm 0.01$ & $0.30 \pm 0.04$ & $0.16 \pm 0.02$ & $0.55 \pm 0.03$ & $0.24 \pm 0.01$ & $0.76 \pm 0.10$ & $0.29 \pm 0.04$ \\
    \textit{Square} & 0 & $0.50 \pm 0.03$ & $0.22 \pm 0.01$ & $0.29 \pm 0.04$ & $0.16 \pm 0.02$ & $0.53 \pm 0.03$ & $0.23 \pm 0.01$ & $0.85 \pm 0.09$ & $0.33 \pm 0.03$ \\
    \textit{Frequency} & 0 & $0.51 \pm 0.03$ & $0.23 \pm 0.01$ & $0.35 \pm 0.05$ & $0.19 \pm 0.03$ & $0.54 \pm 0.03$ & $0.24 \pm 0.01$ & $0.76 \pm 0.10$ & $0.29 \pm 0.04$ \\
    \hline
    \textit{Time} & 1 & $0.45 \pm 0.02$ & $0.20 \pm 0.01$ & $0.26 \pm 0.03$ & $\mathbf{0.14 \pm 0.02}$ & $0.49 \pm 0.03$ & $0.21 \pm 0.01$ & $0.70 \pm 0.08$ & $0.27 \pm 0.03$ \\
    \textit{Square} & 1 & $\mathbf{0.42 \pm 0.02}$ & $\mathbf{0.19 \pm 0.01}$ & $0.33 \pm 0.04$ & $0.18 \pm 0.02$ & $\mathbf{0.42 \pm 0.03}$ & $\mathbf{0.18 \pm 0.01}$ & $0.62 \pm 0.09$ & $0.24 \pm 0.03$ \\
    \textit{Frequency} & 1 & $0.46 \pm 0.02$ & $0.20 \pm 0.01$ & $0.30 \pm 0.04$ & $0.16 \pm 0.02$ & $0.48 \pm 0.03$ & $0.21 \pm 0.01$ & $0.69 \pm 0.08$ & $0.26 \pm 0.03$ \\
    \hline
    \textit{Time} & 2 & $0.46 \pm 0.02$ & $0.21 \pm 0.01$ & $\mathbf{0.27 \pm 0.03}$ & $0.15 \pm 0.02$ & $0.49 \pm 0.03$ & $0.22 \pm 0.01$ & $0.69 \pm 0.09$ & $0.26 \pm 0.03$ \\
    \textit{Square} & 2 & $0.50 \pm 0.02$ & $0.22 \pm 0.01$ & $0.34 \pm 0.04$ & $0.19 \pm 0.02$ & $0.51 \pm 0.03$ & $0.23 \pm 0.01$ & $0.79 \pm 0.09$ & $0.30 \pm 0.03$ \\
    \textit{Freq} w \slash $|\mathrm{STFT}|$  & 2 & $0.46 \pm 0.02$ & $0.21 \pm 0.01$ & $0.28 \pm 0.03$ & $0.15 \pm 0.02$ & $0.49 \pm 0.03$ & $0.21 \pm 0.01$ & $0.71 \pm 0.09$ & $0.27 \pm 0.03$ \\
    \textit{Freq} w \slash sinus and cosinus & 2 & $0.46 \pm 0.02$ & $0.20 \pm 0.01$ & $0.28 \pm 0.03$ & $0.16 \pm 0.02$ & $0.48 \pm 0.03$ & $0.21 \pm 0.01$ & $0.74 \pm 0.09$ & $0.28 \pm 0.03$ \\
    \rowcolor{Gray} \textit{Frequency} & 2 & $\mathbf{0.42 \pm 0.02}$ & $\mathbf{0.19 \pm 0.01}$ & $0.33 \pm 0.05$ & $0.18 \pm 0.03$ & $0.43 \pm 0.03$ & $0.19 \pm 0.01$ & $\mathbf{0.55 \pm 0.09}$ & $\mathbf{0.21 \pm 0.04}$ \\
    \hline \hline
    \end{tabular}
}
\end{table*}

\begin{table*}[ht!]
\caption{Ablation study of attention map using frequency kernels on synthetic data with clean speech. Gray row highlights the proposed approach.}
\label{tab:ablationAttention}
    \centering
        \adjustbox{max width=\textwidth}{%
    \begin{tabular}{l|cc|cc|cc|cc|cc}
    \hline \hline
        \multirow{2}*{Attention} & \multicolumn{2}{c|}{Average} & \multicolumn{2}{c|}{$[1,2)$} & \multicolumn{2}{c|}{$[2,4)$} & \multicolumn{2}{c|}{$[4, 8)$} & \multicolumn{2}{c}{$[8, 14)$} \\
     & $\mathcal{L}_1$ & $\mathrm{r} \mathcal{L}_1$ & $\mathcal{L}_1$ & $\mathrm{r} \mathcal{L}_1$ & $\mathcal{L}_1$ & $\mathrm{r} \mathcal{L}_1$ & $\mathcal{L}_1$ & $\mathrm{r} \mathcal{L}_1$ & $\mathcal{L}_1$ & $\mathrm{r} \mathcal{L}_1$ \\
     \hline
     None & $0.14 \pm 0.01$ & $0.05 \pm 0.00$ & $0.13 \pm 0.01$ & $0.09 \pm 0.01$ & $0.12 \pm 0.01$ & $0.04 \pm 0.00$ & $0.15 \pm 0.01$ & $0.03 \pm 0.00$ & $0.28 \pm 0.05$ & $0.03 \pm 0.00$ \\
     on spectrogram & $0.12 \pm 0.00$ & $\mathbf{0.04 \pm 0.00}$ & $\mathbf{0.12 \pm 0.01}$ & $\mathbf{0.08 \pm 0.01}$ & $\mathbf{0.10 \pm 0.01}$ & $0.04 \pm 0.00$ & $0.13 \pm 0.01$ & $\mathbf{0.02 \pm 0.00}$ & $0.22 \pm 0.03$ & $\mathbf{0.02 \pm 0.00}$ \\
     \rowcolor{Gray} \textbf{on everything} & $\mathbf{0.11 \pm 0.00}$ & $\mathbf{0.04 \pm 0.00}$ & $\mathbf{0.12 \pm 0.01}$ & $\mathbf{0.08 \pm 0.01}$ & $\mathbf{0.10 \pm 0.00}$ & $\mathbf{0.03 \pm 0.00}$ & $\mathbf{0.11 \pm 0.01}$ & $\mathbf{0.02 \pm 0.00}$ & $\mathbf{0.16 \pm 0.02}$ & $\mathbf{0.02 \pm 0.00}$ \\
    \hline \hline
    \end{tabular}
}
\end{table*}

\begin{table}[ht!]
\caption{Ablation study of attention map using frequency kernels on hybrid and real data. Gray row highlights the proposed approach.}
\label{tab:ablationAttentionHybridReal}
    \centering
        \adjustbox{max width=0.49\textwidth}{%
    \begin{tabular}{l|cc|cc|cc}
    \hline \hline
        \multirow{2}*{Attention} & \multicolumn{2}{c|}{QMULTIMIT} & \multicolumn{2}{c|}{VoiceHome~-~2} & \multicolumn{2}{c}{STARSS22} \\
     & $\mathcal{L}_1$ & $\mathrm{r} \mathcal{L}_1$ & $\mathcal{L}_1$ & $\mathrm{r} \mathcal{L}_1$ & $\mathcal{L}_1$ & $\mathrm{r} \mathcal{L}_1$ \\
     \hline
     None & $2.01 \pm 0.06$ & $0.21 \pm 0.01$ & $0.78 \pm 0.09$ & $0.40 \pm 0.06$ & $0.45 \pm 0.02$ & $0.20 \pm 0.01$ \\
     on spectrogram & $\mathbf{1.87 \pm 0.06}$ & $\mathbf{0.19 \pm 0.01}$  & $0.73 \pm 0.10$ & $0.36 \pm 0.06$ & $0.45 \pm 0.02$ & $0.20 \pm 0.01$\\
     \rowcolor{Gray} \textbf{on everything} &  $1.90 \pm 0.06$ & $0.20 \pm 0.01$ & $\mathbf{0.63 \pm 0.08}$ & $\mathbf{0.32 \pm 0.05}$ & $\mathbf{0.42 \pm 0.02}$ & $\mathbf{0.19 \pm 0.01}$\\
    \hline \hline
    \end{tabular}
}
\end{table}

\subsection{Cross-corpus generalization}

Tests have been carried out in a cross-corpus training-testing setup, e.g., synthetic-hybrid, synthetic-real, hybrid-real, VoiceHome-STARSS. The model yields very large errors in case no finetuning is performed, as it can be inspected in Table~\ref{tab:cross_dataset_noFT}. This behavior highlights the discrepancy of feature patterns among different acoustic scenarios, levels of acoustical realism, and different distance distributions. If the model is fine-tuned to  a different realistic scenario, the performance is slightly worse that the case when the model starts with random weights. The results of this situation is shown in Table~\ref{tab:cross_dataset_FT}.

\begin{table}[ht!]
\caption{Cross-dataset generalization tests without finetuning.}
\label{tab:cross_dataset_noFT}
    \centering
        \adjustbox{max width=\textwidth}{%
        \begin{tabular}{ cc|c|c|c } 
        \multicolumn{5}{c}{\textbf{Test w/o finetuning}} \\ 
        \multirow{5}{*}{\rotatebox{90}{\textbf{Training}}} & & \textbf{Synthetic} & \textbf{Hybrid} & \textbf{Real} \\ \cline{2-5}
            & \textbf{Synthetic} & $0.11 \pm 0.00$ & $4.28 \pm 0.45$ & $4.14 \pm 0.08$ \\ \cline{2-5}
            & \textbf{Hybrid} & $6.80 \pm 0.59 $ & $1.52 \pm 0.12$ & $3.76 \pm 0.56$ \\ \cline{2-5}
            & \textbf{Real} & $2.26 \pm 0.38$ & $8.22 \pm 0.54$ & $0.42 \pm 0.02$ \\ 
        \end{tabular}
}
\end{table}

\begin{table}[ht!]
\caption{Cross-dataset generalization tests with finetuning.}
\label{tab:cross_dataset_FT}
    \centering
        \adjustbox{max width=\textwidth}{%
        \begin{tabular}{ cc|c|c|c } 
        \multicolumn{5}{c}{\textbf{Test w/ finetuning}} \\ 
        \multirow{5}{*}{\rotatebox{90}{\textbf{Training}}} & & \textbf{Synthetic} & \textbf{Hybrid} & \textbf{Real} \\ \cline{2-5}
        & \textbf{Synthetic} & $0.11 \pm 0.00$ & $1.57 \pm 0.23$ & $0.47 \pm 0.05$ \\ \cline{2-5}
        & \textbf{Hybrid} & $0.18 \pm 0.04$ & $1.52 \pm 0.12$ & $0.45 \pm 0.05$ \\ \cline{2-5}
        & \textbf{Real} & $0.11 \pm 0.02$ & $1.54 \pm 0.22 $ & $0.42 \pm 0.02$ \\ 
        \end{tabular}
}
\end{table}

\section{Discussion}\label{sec:discussion}

From the results of the noisy scenario in the synthetic dataset, it is important to highlight that even a minimal amount of noise severely corrupts phase-based features, which have been identified as the most critical information in our analysis of clean speech. For instance, the presence of direct sound and echo patterns, characterized by transients in the clean signal, becomes blurred over time due to the presence of noise and late reverberation, resulting in a loss of phase coherence across frequencies. This behavior, however, does not occur in the hybrid dataset where the effect of high \ac{SNR} in the recordings does not correspond to a similar increase in estimation performance. That may be due to the recordings of the \acp{RIR} having a level of inherent measurement noise, which limits the effective \ac{SNR} that we can achieve in the hybrid simulations.

The imposition of the loss in~\eqref{eq:loss} is required for predicting a time-wise distance vector. Due to the lack of baselines and datasets in the literature, only a single value of distance of the sound source is assigned for each time bin to ease the distance tracking task. Generally, this characteristic in audio datasets is referred as \textit{weak labels}~\cite{Kumar_2017_IJCNN}. Without time-wise distance references, denoted as \textit{strong labels}, the model encounters challenges in fine-tuning its predictions, decreasing its overall performance. This scenario has been studied in literature for tasks that require a fine temporal resolution output, such as \ac{SED}~\cite{Neri_2022_Access} and \ac{SELD}~\cite{Martin_2023_TASLP}.

Furthermore, it is important to acknowledge that certain portions of the audio data encompass segments where speech information is absent or indiscernible. Consequently, this scarcity of informative speech content can considerably undermine the effectiveness and reliability of the predictors.

In this direction, the proposed attention module can improve the ability of the model (Tab.~\ref{tab:ablationAttentionHybridReal}) to identify the speech information that is relevant for the estimation of the distance. However, it is important to note that the attention module is learned by the model itself, without any direct supervision.

To address these limitations, a potential avenue for improvement emerges, centering around the generation of more comprehensive and fine-grained labels. By augmenting the dataset with \textit{strong labels} that introduces both speech activity and speaker distance estimation, the model may acquire a better understanding of the room acoustics. In addition, this augmentation enables the model to leverage additional contextual cues and refine its predictions, enhancing its performance in accurately estimating speaker distances and capturing the dynamics of speech activity.

Moreover, one of the key areas for improvement is the availability of larger datasets of real recordings with a greater number of rooms and various speaker-microphone configurations. A larger dataset would enable the model to learn more diverse and representative acoustic characteristics, leading to improved performance in distance estimation tasks. Moreover, it could also improve the generalization ability of the approach, as it has been demonstrated how the performance of the proposed model is dependent on the nature of the audio recording (synthetic, hybrid or real). Additionally, by including different room types and microphone placements, the model can better generalize across various real-world scenarios. Furthermore, the use of a transformer-based~\cite{vaswani2017attention} approach could be explored, leveraging a larger amount of data. Transformer models have shown remarkable success in various natural language processing tasks and have the potential to capture complex patterns and dependencies in acoustic data. Exploiting transformer architectures could enhance the model's ability to estimate distances accurately.

Another possibility for future research is the integration of time-wise distance ground truth, as previously mentioned in the discussion section. By considering temporal information in addition to spatial cues, the model could potentially estimate the distance of a sound source more accurately. This would provide valuable insights in scenarios where multiple sound sources are present. Estimating and tracking the distance of a moving source is an application of interest that is scarcely explored in the literature.

\section{Conclusions}\label{sec:conclusion}
This work has explored the task of speaker distance estimation in noisy and reverberant environments. Multiple configurations, in terms of kernel size and recurrent layers of the model, have been provided, motivating the proposed architecture. In fact, the use of rectangular filters across the frequency dimension and the presence of \acp{GRU} layers yields the best performance in terms of distance errors. The experimental results obtained from the proposed model have demonstrated remarkable precision in scenarios where several types of \acp{RIR} are employed. In a noiseless synthetic scenario where \acp{RIR} have been generated with a room-source simulator, the model has achieved an absolute error of only $0.11$ meters. With recorded \acp{RIR}, an absolute error of about $1.30$ meters has been obtained. In the real scenario with on-field recordings, where unpredictable environmental factors and noise were prevalent, the model yielded an absolute error of approximately $0.50$ meters. These results underscore the model's resilience and its capacity to effectively manage various realistic scenarios. Variations in performance across these scenarios can be attributed to differences in the distribution of acoustic parameters, such as the distance from the sound source. Analysis on moving sound sources in single-channel recordings will be carried out as a future work.

\bibliographystyle{IEEEbib}
\bibliography{bibs}

\end{document}